\documentclass[aps,prl,twocolumn,superscriptaddress]{revtex4-2}
\usepackage{amsmath,amssymb,bm,amsthm,amsfonts,graphicx}
\usepackage{xspace}
\usepackage[dvipsnames]{xcolor}         
\usepackage{fancyhdr}                          
\usepackage{newtxtext}                        
\usepackage{color}
\usepackage{multirow}
\usepackage[bookmarks=false,linkcolor=blue,urlcolor=blue,colorlinks,citecolor=blue]{hyperref}
\usepackage{gensymb}
\usepackage{mhchem}    
\newcommand{\beq}    {\begin{equation}}
\newcommand{\enq}    {\end{equation}}
\newcommand{\ceq}[1] {(\ref{#1})}

\newcommand{\bb}     {{\bf b}}

\newcommand{\kk}     {{\bf k}}

\DeclareMathOperator {\sgn}{sgn}

\newcommand{\cdas}   {${\rm Cd_3As_2}$\xspace}

\newcommand{\ssigma}     {{\boldsymbol\sigma}}

\begin{document}

\title{Spontaneous Time-Reversal-Symmetry-Breaking State in Proximity-Induced Superconductivity}

\author{Jonathan Schirmer}
\affiliation{Department of Physics, William \& Mary, Williamsburg, Virginia 23187, USA}
\author{Wei Pan}
\affiliation{Sandia National Laboratories, Livermore, California 94551, USA}
\author{Javad Shabani}
\affiliation{Center for Quantum Information Physics, New York University, New York, New York 10003, USA}
\author{Enrico Rossi}
\affiliation{Department of Physics, William \& Mary, Williamsburg, Virginia 23187, USA}

\date{\today}

\begin{abstract}
Recent experiments on superconducting heterostructures formed by an $s$-wave superconductor and a Dirac semimetal
suggest that in these systems the superconducting state can break time-reversal symmetry
even though no magnetic order or unconventional pairing is present in any of the structure's components.
We present a minimal mechanism for a time-reversal-symmetry-breaking superconducting state
in multiband systems proximity-coupled to a conventional $s$-wave superconductor.
We derive the conditions for this state, find that it is the ground state wherever it exists,
and obtain the phase diagrams. We find that in some situations the phase diagrams describe
reentrant behavior: by continuously varying a parameter the system can undergo multiple
transitions between a time-reversal-symmetric state and a state that spontaneously breaks time-reversal symmetry.
The results help identify heterostructures that can be engineered
to obtain nonreciprocal superconducting transport effects even when no magnetic field or magnetization is present.
\end{abstract}

\maketitle

%
%
The most common superconductors couple electrons in states that are
time-reversal pairs. As a consequence, such superconductors are invariant
under time reversal. Superconducting states that break time-reversal
symmetry (TRS) are very rare and extremely interesting~\cite{sigrist1991,sigrist1998},
both from a fundamental point of view and for possible technological applications.
For instance, such states can give rise to a superconducting diode effect 
(SDE)~\cite{Ando2020,Daido2022,yuan2022,he2022,Nadeem2023,hou2023,shaffer2025},
an effect that can be used to realize ideal, dissipationless diodes to perfectly isolate quantum 
devices, and some of these states are topological~\cite{qi2011a,alicea2012,sato2017,ryu2024},
supporting non-Abelian fermionic states that can be used to realize fault-tolerant, topologically
protected qubits~\cite{kitaev2003,nayak2008,sarma2015}.
Any microscopic mechanism that can give rise to a TRS-breaking (TRSB) superconducting 
state is therefore of great interest.

One way to obtain TRSB superconducting states is by applying an external magnetic field to a superconductor with strong
spin-orbit coupling. This is the approach used in superconductor-semiconductor heterostructures designed
to realize Majorana bound states~\cite{kitaev2001,sau2010a,lutchyn2010c,oreg2010,mourik2012,lutchyn2018,dassarma2023}.
In some materials, for instance FeTeSe, superconductivity
appears to coexist with ferromagnetism, resulting in a superconducting state with broken TRS~\cite{zaki2021,qiu2023,liu2026}.
In these examples TRS is explicitly broken, either by applying
a magnetic field, or by the coexistence of ferromagnetism.
Recent experiments on asymmetric superconducting quantum interference devices (SQUIDs)
based on Al/\cdas heterostructures~\cite{schumann2016,lygo2023} have observed
a Josephson diode effect (JDE)~\cite{yu2024,travaglini2025} 
in the absence of any magnetic field.
This is surprising given that 
the breaking of TRS is a necessary condition for 
the presence of the JDE~\cite{misaki2021,jeon2022,kokkeler2022,lin2022,ilic2022,wu2022,qu2024,yerin2024,yerin2025}, 
none of the individual components forming the heterostructure are ferromagnetic,
or are expected to be in a ground state that otherwise breaks TRS, 
and no magnetic impurities are present. 
The measurements reported in Ref.~\cite{yu2024,travaglini2025} therefore
suggest that in Al/\cdas heterostructures we can have a superconducting ground state 
spontaneously breaking TRS with no net magnetization.
Using a phenomenological Ginzburg-Landau model, Ref.~\cite{NgNagaosa2009}
suggested that in a Josephson junction formed by a two-gap superconductor 
and a conventional $s$-wave superconductor, a TRSB superconducting state can be
realized even when no magnetic field, or magnetic order, explicitly breaks TRS.
However, the model of Ref.~\cite{NgNagaosa2009} did not provide insight
into the microscopic conditions for the establishment of a TRSB superconducting state.
A closely related mechanism for a three-gap superconductor
was presented in Ref.~\onlinecite{StanevTesanovic2010}, in which
all three superconducting components are
intrinsic, self-consistent order parameters of the same material, 
and interact repulsively with each other via pair-scattering terms.
In general, in a superconductor with three or more superconducting pairings, 
when the interactions among pairings are frustrated, the
relative phases cannot all settle at $0$ or $\pi$, and the order parameters
cant into a complex configuration that breaks TRS and is degenerate with its time
reverse~\cite{StanevTesanovic2010,TanakaYanagisawa2010,Garaud2011,MaitiChubukov2013}.
The presence of multiple intrinsic pairings, and frustrated repulsive
interactions, places stringent demands on a material's electronic properties, a fact that explains
why the observation of intrinsic TRSB superconducting states is expected to be rare
and to this day is still not unambiguously established.

In this work we show how a TRSB superconducting state,
with no net magnetization and only conventional $s$-wave pairings, can arise
in heterostructures in which a conventional $s$-wave superconductor, such as Al,
is coupled, via the proximity effect, to a multiband metal with residual repulsive interband
interactions. 
Unlike intrinsic multiband superconductors, in the system we consider the frustration
is extrinsic: the proximity effect favors a 0 relative phase difference ($\varphi_{12}$) between
the pairings in the metal, whereas the repulsive interband interactions favor a $\pi$
relative phase difference.
Using a microscopic model, we obtain the general conditions for the establishment
of a TRSB superconducting state in the absence of any magnetic field or ordering.
In addition, we find that in the presence of asymmetries between the two bands
the proximity setup can lead to a phase diagram in parameter space exhibiting
reentrant behavior: as one of the parameters is varied, for instance the average induced
gap for the two bands, the system can undergo first a transition from a time-reversal
symmetric state to a TRSB state, and then a second transition from the TRSB state to
a time-reversal symmetric state. 
We then discuss the relevance of our results
for heterostructures formed by an $s$-wave superconductor in proximity to a two-dimensional electron gas (2DEG)
with strong spin-orbit coupling (SOC), or a Dirac semimetal
such as \cdas, and for Josephson junctions (JJs) based on these structures.

\begin{figure}[htb]
  \begin{center}
    \centering
    \includegraphics[width=\linewidth]{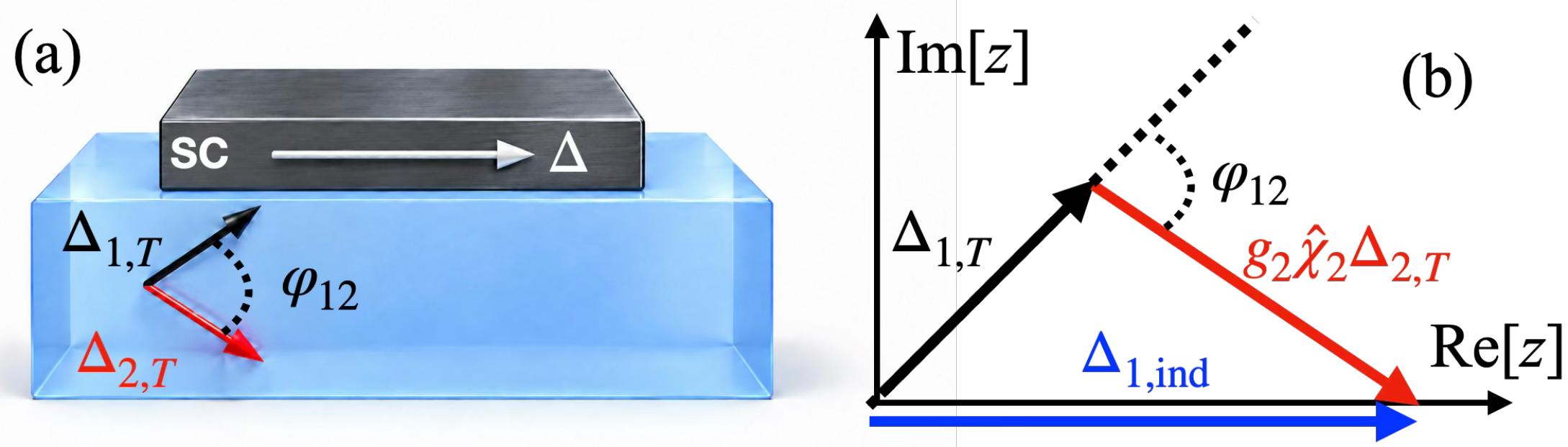}
    \caption{(a) Sketch of a heterostructure formed by a superconductor, shown in gray, 
                 placed on a 2-band metal in which two superconducting pairings are induced.
             (b) Geometric construction showing schematically the solution of the first gap
                 equation in Eq.~\ceq{eq.gap}, from which follows the triangle inequality that must
                 be satisfied to have $\varphi_{12}\neq 0, \pi$.}
    \label{fig.sketch}
  \end{center}
\end{figure}

The system considered, shown schematically in Fig.~\ref{fig.sketch}~(a), is described by the 
Hamiltonian:
$ H = H_{SC} + H_{M} + H_{\rm int} + H_T$,
where $H_{SC}$ is the Hamiltonian for the conventional $s$-wave superconductor,
$H_M$ and $H_{\rm int}$ are the kinetic and interaction parts, respectively, 
of the metal's Hamiltonian, and $H_T$ is the operator describing tunneling processes
between the superconductor and the metal.
We assume the metal to have two bands that we denote with indices $j=1,2$,
and describe the superconductor via the Bogoliubov-de Gennes (BdG) approach
with $\Delta$ the superconductor's pairing field that, to very good approximation
for superconductors like Al, can be taken to be momentum independent.
After integrating out the superconductor's degrees of freedom,
the effect of $H_{SC}$, mediated by $H_T$, on the metal 
is to induce pairing fields $\Delta_{j,\rm ind}$
in the two bands.
The effective Hamiltonian for the metal then takes the form
\begin{align}
 H_{\rm eff}\!\! =\!\! & \sum_{\kk\sigma j}(\epsilon_{\kk \sigma j}-\mu)c^\dagger_{\kk\sigma j}c_{\kk\sigma j} \nonumber \\
                     & -\sum_{\kk j}[\Delta_{j,\rm ind}c^\dagger_{\kk\uparrow j}c^\dagger_{-\kk\downarrow j}+ h.c.]  + H_{\rm int}
 \label{eq.Heff}
\end{align}
where $c^\dagger_{\kk\sigma j}$ and $c_{\kk\sigma j}$ are the creation and annihilation operators,
respectively, for an electron with momentum $\kk$, spin $\sigma$, in band $j$,
$\epsilon_{\kk\sigma j}$ is the corresponding energy, with
$\epsilon_{\kk\uparrow j}=\epsilon_{-\kk\downarrow j}\equiv\epsilon_{\kk, j}$ by time-reversal symmetry,
and $\mu$ is the chemical potential.
We assume that the repulsive interaction between Cooper pairs in different bands
is the dominant one, so that
\beq
 H_{\rm int} = \frac{U}{V}\sum_{\kk,\kk'}[c^\dagger_{\kk\uparrow 1}c^\dagger_{-\kk\downarrow 1}
                                c_{-\kk'\downarrow 2}c_{\kk'\uparrow 2} + h.c.]
 \label{eq.Hint}
\enq
where $U>0$ is the interaction strength, and $V$ is the volume of the system.
Let $\Delta_j\equiv (U/V)\sum_{\kk}\langle c_{-\kk\downarrow j}c_{\kk\uparrow j}\rangle$
so that within the mean-field approximation $H_{\rm int}$ becomes
\begin{align}
 H_{\rm int}^{MF} = & \sum_{\kk}[\Delta_2 
 c^\dagger_{\kk\uparrow 1}c^\dagger_{-\kk\downarrow 1}
 +\Delta_1 c^\dagger_{\kk\uparrow 2}c^\dagger_{-\kk\downarrow 2} + h.c.] \nonumber \\
 & -\frac{V}{U}(\Delta_1^*\Delta_2 + \Delta_1\Delta_2^*).
 \label{eq.HintMF}
\end{align}
We assume bands 1 and 2 to be well separated, namely,
the energy separation between the two bands to be much larger than
the superconducting gap, a condition that is relevant for the systems of interest.

The total pairing gap in band $j$, $\Delta_{j,T}=\Delta_{j,\rm ind}-\Delta_{\bar j}$ ($\bar j\neq j$), 
is the sum of $\Delta_{j,\rm ind}$ and the one
induced by $H_{\rm int}$, given by the pairing field $\Delta_{\bar j}$ of the other band. 
We have $\Delta_{j} = U \Delta_{j,T}\chi_j(|\Delta_{j,T}|)$,  where
$\chi_j(|\Delta_{j,T}|)=(1/V)\sum_{\kk} \tanh (\beta E_{\kk,j}/2)/(2E_{\kk,j})$
is the pairing susceptibility  with  
$E_{\kk\sigma j}=\sqrt{\xi_{\kk j}^2 + |\Delta_{j,T}|^2}$,
$\xi_{\kk,j}=(\epsilon_{\kk,j}-\mu)$,
and $\beta=1/k_BT$ the inverse temperature.
In the remainder we consider the zero-temperature limit, $T=0$. 
Assuming the density of states in each band to be approximately 
constant around the Fermi energy, and the cutoff energy $\omega_c$ 
used as the upper limit in the integral to obtain $\chi_j$ to be much larger than $|\Delta_{j,T}|$, 
we obtain $\chi_j\approx N_j(0)\ln(2\omega_c/|\Delta_{j,T}|)$,
where $N_j(0)$ is the density of states at the Fermi energy in band $j$.
Setting $g_j\equiv U N_j(0)$ and $\hat\chi_j\equiv \chi_j/N_j(0)\approx \ln(2\omega_c/|\Delta_{j,T}|)$,
the pairing gaps are given by the coupled self-consistent equations:
\begin{equation}
\begin{pmatrix}
1                            & g_2\hat\chi_2 (|\Delta_{2,T}|)\\
g_1\hat\chi_1(|\Delta_{1,T}|) & 1
\end{pmatrix}
\begin{pmatrix}
\Delta_{1,T} \\
\Delta_{2,T}
\end{pmatrix}
=
\begin{pmatrix}
\Delta_{1,\rm ind} \\
\Delta_{2,\rm ind}
\end{pmatrix}
\label{eq.gap}
\end{equation}

In general, $\Delta_{j,T}$ are complex numbers. We can introduce
the phases $\varphi_j$ ($j=1,2$) such that $\Delta_{j,T}=|\Delta_{j,T}|e^{i\varphi_j}$.
We see that the system of equations~\ceq{eq.gap} is characterized by frustration:
the induced gaps, both real, tend to align $\Delta_{1,T}$ 
and $\Delta_{2,T}$, whereas the interaction $U$ tends to anti-align them.
When $g_1\hat\chi_1 g_2\hat\chi_2\neq 1$, Eqs.~\ceq{eq.gap} only admit solutions
in which $\Delta_{j,T}$ are either aligned, $\varphi_{12}\equiv\varphi_1-\varphi_2=0$, or anti-aligned, $\varphi_{12}=\pi$. Both solutions satisfy TRS.
Given that $\Delta_{j,\rm ind}$ are real, a necessary
condition for the imaginary part of Eq.~\ceq{eq.gap} to have nonzero
solutions is that the determinant of the matrix ($M$) in Eq.~\ceq{eq.gap} 
multiplying $(\Delta_{1,T}$, $\Delta_{2,T})^T$ be zero, i.e., 
\beq
 g_1\hat\chi_1(|\Delta_{1,T}|) g_2\hat\chi_2(|\Delta_{2,T}|) = 1 
 \label{eq.det}
\enq
In this case, the Leggett mode describing fluctuations of $\varphi_{12}$
becomes soft and solutions with $\varphi_{12}\neq 0,\pi$ are possible. These solutions
describe TRSB superconducting states.

When $\det(M)=0$, in order to have nontrivial solutions,
the real part of the 
system of equations~\ceq{eq.gap}  must satisfy the rank consistency condition
${\rm rank}[M]={\rm rank}[M|\vec{\Delta}_{\rm ind}]$ where 
$\vec{\Delta}_{\rm ind}\equiv (\Delta_{1,\rm ind},\Delta_{2,\rm ind})^T$, namely,
\beq
 g_1\hat\chi_1(|\Delta_{1,T}|)=\Delta_{2,\rm ind}/\Delta_{1,\rm ind}.
 \label{eq.comp}
\enq
Equations~\ceq{eq.det},~\ceq{eq.comp}, combined with 
the first (or, equivalently, the second) equation in~\ceq{eq.gap}, which can be rewritten as
\beq
 |\Delta_{1,T}|e^{i\varphi_1}+g_2\hat\chi_2|\Delta_{2,T}|e^{i\varphi_2}=\Delta_{1,\mathrm{ind}}
\label{eq.re-im}
\enq
completely define $|\Delta_{1,T}|$, $|\Delta_{2,T}|$ and $\varphi_{12}$.

Let $r_{12}\equiv\Delta_{1,\rm ind}/\Delta_{2,\rm ind}$.
Using Eqs.~\ceq{eq.det},~\ceq{eq.comp}, for $\Delta_{1,\rm ind}\neq 0$,
$\Delta_{2,\rm ind}\neq 0$, and $|\Delta_{j,T}|\ll \omega_c$, we find
\beq
 |\Delta_{1,T}| = 2\omega_c e^{-1/(g_1 r_{12})}; \hspace{0.5cm}
 |\Delta_{2,T}| = 2\omega_c e^{-r_{12}/g_2}.
 \label{eq.abs_deltas}
\enq 
Taking the modulus squared of Eq.~\ceq{eq.re-im} we find
\beq
 |\Delta_{1,T}|^{2}\!\!+\!g_2^2\hat\chi_2^{2}|\Delta_{2,T}|^{2}
 \!\!+\!\!2g_2\hat\chi_2|\Delta_{1,T}||\Delta_{2,T}|\!\cos\varphi_{12}\!\!=
 \!\!\Delta_{1, {\rm ind}}^2
 \label{eq.modsq}
\enq
and therefore 
\beq
 { \varphi_{12} = \pm\arccos\!\left[ \frac{ \Delta_{1,\mathrm{ind}}^{\,2} -|\Delta_{1,T}|^{2} -(g_2\hat\chi_2)^{2}|\Delta_{2,T}|^{2} }{ 2g_2\hat\chi_2|\Delta_{1,T}||\Delta_{2,T}| } \right]} 
 \label{eq.phi12}
\enq
where the two solutions with opposite $\varphi_{12}$ are related by TRS.
In the remainder we use $\omega_c$ as our energy scale. Setting $\omega_c=1$
and using Eqs.~\ceq{eq.abs_deltas} and~\ceq{eq.phi12}, we finally find:
\beq
 \varphi_{12} = \pm\arccos\!\left[ \frac{ \hat\Delta_{1,\mathrm{ind}}^{\,2} 
  -4e^{-2/(g_1r_{12})} -4r_{12}^{\,2}e^{-2r_{12}/g_2} }{ 8r_{12}\, 
  e^{-\left[\frac{1}{g_1r_{12}}+\frac{r_{12}}{g_2}\right]} } \right]
 \label{eq.phi12b}
\enq
where $\hat\Delta_{1, \mathrm{ind}}\equiv \Delta_{1, \mathrm{ind}}/\omega_c$.

Equation~\ceq{eq.phi12b} gives the canting angle $\varphi_{12}$ in terms of
the coupling constants $g_1$, $g_2$, and the ratio $r_{12}$ between the induced gaps.
In particular, the boundary of the region in parameter space 
between the TRS state with $\varphi_{12}=0$ and a TRSB state 
can be obtained by setting the argument of the $\arccos$ in Eq.~\ceq{eq.phi12b} 
equal to $+1$. Similarly, the boundary between the TRS state with  $\varphi_{12}=\pi$
and the TRSB state is obtained by setting the argument of the $\arccos$ equal to $-1$.
The condition for which $\varphi_{12}\neq 0,\pi$ and a TRSB state is present can be determined
by the geometric representation of Eq.~\ceq{eq.re-im} shown in Fig.~\ref{fig.sketch}~(b).
From this figure we see that $\varphi_{12}$ will be different from zero and $\pi$ when
the triangle inequality
\beq
 ||\Delta_{1,T}| - g_2\hat\chi_2|\Delta_{2,T}|| 
 < |\Delta_{1,\rm ind}| 
 < |\Delta_{1,T}|+ g_2\hat\chi_2|\Delta_{2,T}|
 \label{eq.ineq}
\enq
is satisfied. 
The relative phase is a dynamical degree of freedom, called the Leggett mode. 
Approaching the TRSB transition from a TRS state, the phase-difference 
potential flattens and the Leggett mode softens, going critical at the transition.

For given parameters, Eqs.~\ceq{eq.gap} can have several solutions: one or more
TRS solutions and
the pair of TRSB solutions $\pm\varphi_{12}$.
To identify the ground state, we compare their energies at $T=0$.
The difference $\Delta E\equiv E-E_N$ between the expectation value $E$ of
$H_{\rm eff}$ in the BCS state and the normal-state
energy $E_N$ is, for $|\Delta_{j,T}|\ll\omega_c$,
\begin{align}
 \Delta E=& 
 \frac{1}{U}\sum_{j}g_j\Big[|\Delta_{j,T}|^2\big(\hat\chi_j-\tfrac{1}{2}\big) 
  -2\Delta_{j,\rm ind}\hat\chi_j|\Delta_{j,T}|\cos\varphi_j\Big] \nonumber \\
   &+\frac{2}{U}g_1g_2\hat\chi_1\hat\chi_2|\Delta_{1,T}||\Delta_{2,T}|\cos\varphi_{12}.
 \label{eq.energy}
\end{align}
The three terms are the kinetic-energy cost, the energy gained by aligning
the total pairing fields with the pairing of the parent superconductor, and 
the energy due to the interband repulsion.
Let $\Delta E_{\rm TRS}$ and $\Delta E_{\rm TRSB}$ be the values
of $\Delta E$ for the TRS and TRSB solutions, respectively,
given by Eqs.~\ceq{eq.gap}.
Given that such solutions are stationary points of $\Delta{E}$
($\partial{\Delta E}/\partial|\Delta_{j,T}|=0$ and
$\partial\Delta{E}/\partial\varphi_j=0$ being the real and imaginary parts of
Eqs.~\ceq{eq.gap}, respectively),
from  Eq.~\ceq{eq.energy} we obtain
\begin{align}
 \Delta E_{\rm TRS/TRSB}= &-\tfrac{1}{U}\sum_{j=1,2}\tfrac{1}{2} g_j|\Delta_{j,T}|^2
  -\tfrac{1}{U}\Delta_{1,\rm ind}\Delta_{2,\rm ind} \nonumber \\
 +&\tfrac{1}{U}\big(1-g_1\hat\chi_1g_2\hat\chi_2\big)\,\Delta_{1,T}\Delta_{2,T}
 \label{eq.energy_sol}
\end{align}
Notice that the product on the second line is real for every solution given that for the 
TRS states $\Delta_{1,T}$, $\Delta_{2,T}$ are real, and that for the TRSB states 
$(1-g_1\hat\chi_1g_2\hat\chi_2)=0$.
In the whole parameter range that we have considered 
we find that, wherever the TRSB solution exists, it has a lower energy than all the
TRS solutions. 

\begin{figure}[h]
 \includegraphics[width=\linewidth]{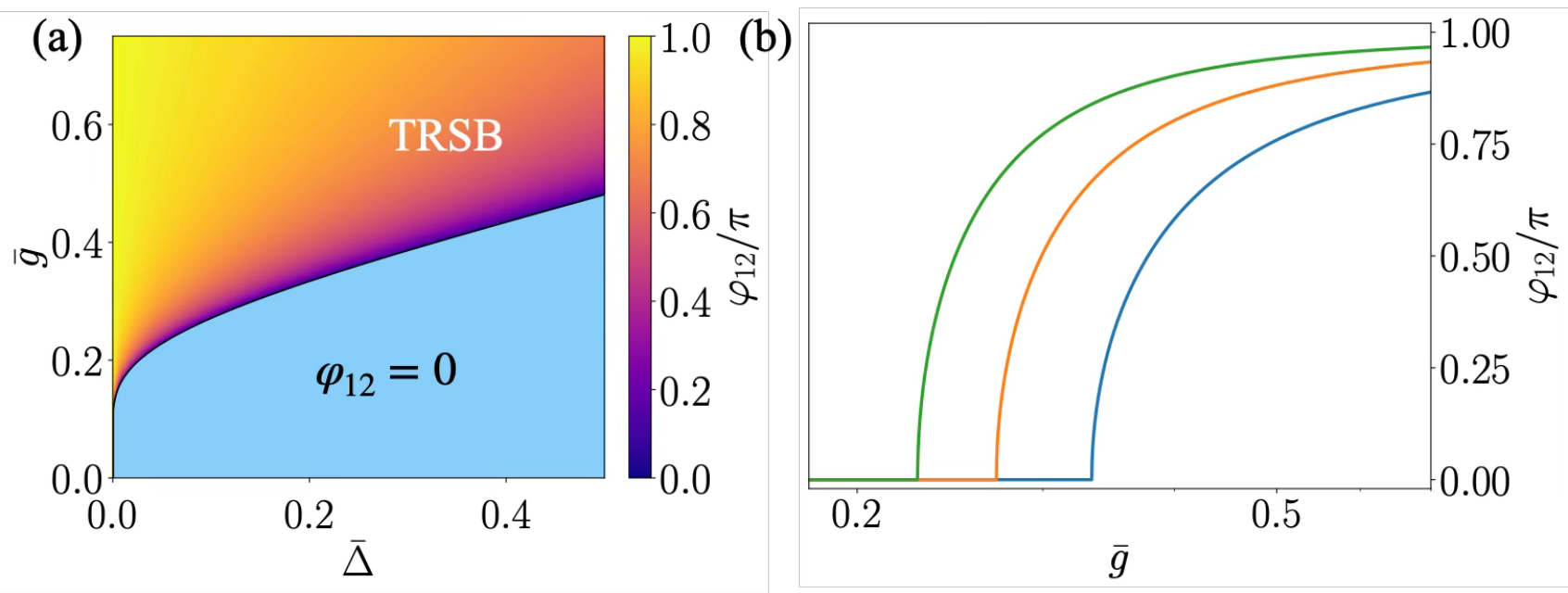}
    \caption{
    Phase diagram in ($\bar\Delta$,$\bar g$) space for the symmetric
    case: $r_{12}=1$, $g_1/g_2=1$.
    (a) $\varphi_{12}$ as a function of $\bar \Delta$ and $\bar g$. 
    (b) Line cuts for constant values of $\bar\Delta$ of panel (a); 
    $\bar\Delta=(0.05, 0.1, 0.2)$ for the green, 
    orange, and blue traces, respectively.
    }
  \label{fig.results_sym}
\end{figure}
Figure~\ref{fig.results_sym}~(a) shows the dependence of $\varphi_{12}$ with respect to
the average induced gap 
$\bar\Delta\equiv(\Delta_{\rm 1, ind}+\Delta_{\rm 2, ind})/2$ and the
average coupling $\bar g\equiv (g_1+g_2)/2$ for the symmetric case
$\Delta_{\rm 1, ind}=\Delta_{\rm 2, ind}$, i.e., $r_{12}=1$, and $g_1=g_2$.
We see that for each value of $\bar\Delta$ there is a critical value, $\bar g_c$, of
$\bar g$ above which the system enters a TRSB phase. 
For the symmetric case, the dependence of $\bar g_c$ on $\bar\Delta$
can be expressed analytically: $\bar g_c=1/\ln(4/\bar\Delta)$.
Fig.~\ref{fig.results_sym}~(a) and
the vertical line cuts presented in 
Fig.~\ref{fig.results_sym}~(b) show that,
for the symmetric case, once the system enters the TRSB phase (for fixed $\bar\Delta$),
the relative phase $\varphi_{12}$ grows with $\bar g$ and for large $\bar g$ saturates to $\pi$.

\begin{figure}[h]
 \includegraphics[width=\linewidth]{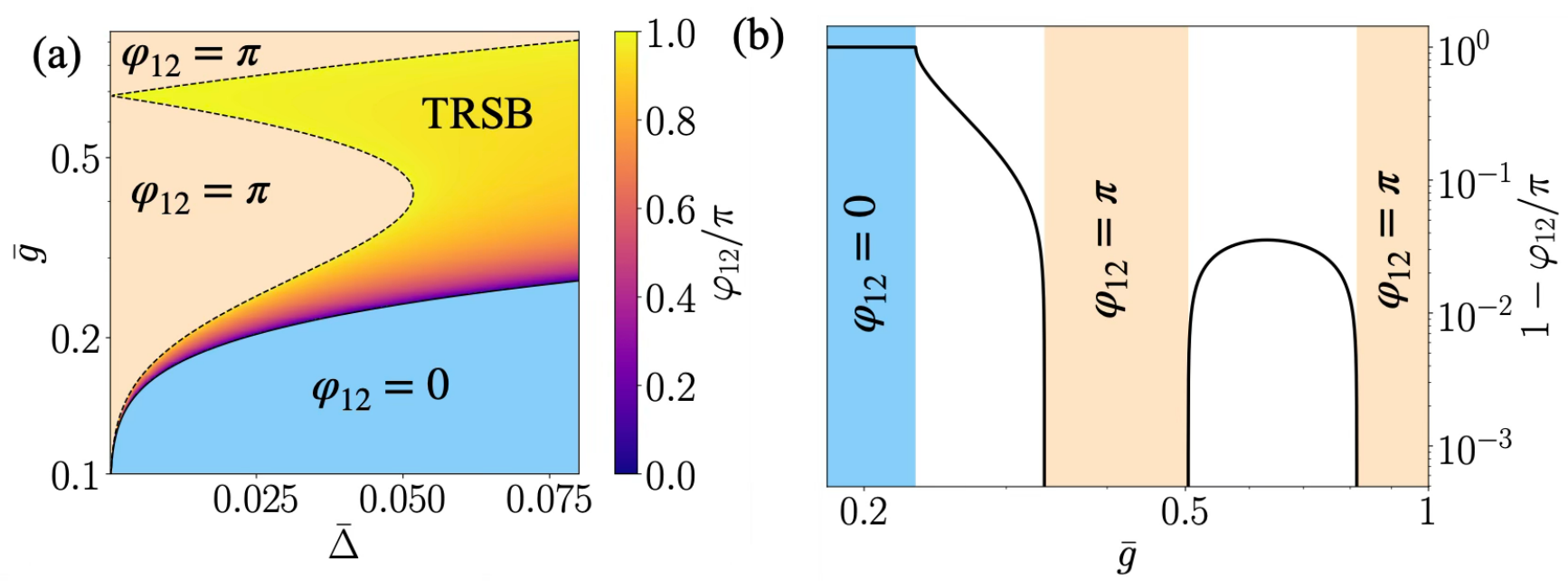}
    \caption{
    Phase diagram and line cut for an asymmetric case: $r_{12}\neq 1$ and $g_1/g_2\neq 1$.
    (a) Phase diagram in ($\bar\Delta$,$\bar g$) space when $r_{12}=0.6$ and $g_1/g_2=2.0$.
    (b) Vertical line cut of panel (a) for $\bar\Delta = 0.045$.
    }
  \label{fig.results_asym}
\end{figure}

For asymmetric cases, $r_{12}\neq 1$ and/or $g_1/g_2\neq 1$, the phase diagram 
can be qualitatively very different from the one obtained for the symmetric case.
For many asymmetric cases the phase diagram exhibits reentrant behavior:
as a parameter is varied the system enters, leaves, and then reenters, the TRSB phase.
This is clearly shown in Fig.~\ref{fig.results_asym}. 
Panel (a) of Fig.~\ref{fig.results_asym} shows the phase diagram as a function
of $\bar\Delta$ and $\bar g$ when $r_{12}=0.6$ and $g_1/g_2=2$. We can see
that there is a range of values of $\bar\Delta$ for which, as $\bar g$ is increased,
the system undergoes multiple phase transitions between a TRS state (with $\varphi_{12}=0$
or $\pi$) and a TRSB state. The line cut of Fig.~\ref{fig.results_asym}~(a) shown in
panel (b) presents the evolution of the deviation from $\pi$ of $\varphi_{12}$ in the TRSB phase between TRS phases.
Additional phase diagrams for asymmetric cases, exhibiting reentrant behavior, 
in the $(g_1/g_2,\bar g)$ and $(g_1/g_2,r_{12})$
parameter spaces, are shown in Fig.~\ref{fig.sm_asym} of the Supplementary Materials (SM).

The phase difference $\varphi_{12}$ is defined modulo $2\pi$;
however, we can distinguish 
two different TRS states with $\varphi_{12}=\pi$: 
the one for which $\Delta_{1,T}$ and $\Delta_{2,T}$
are antiparallel and parallel, respectively, to the induced gaps, i.e., 
$(\varphi_1,\varphi_2)=(\pi,0)$, 
and the one for which $(\varphi_1,\varphi_2)=(0,\pi)$.
In Fig.~\ref{fig.results_asym}, the two $\varphi_{12}=\pi$ TRS phases
separated by a TRSB phase correspond to the two different states
$(\varphi_1,\varphi_2)=(\pi,0)$ and $(\varphi_1,\varphi_2)=(0,\pi)$.

\begin{figure}[h]
 \includegraphics[width=\linewidth]{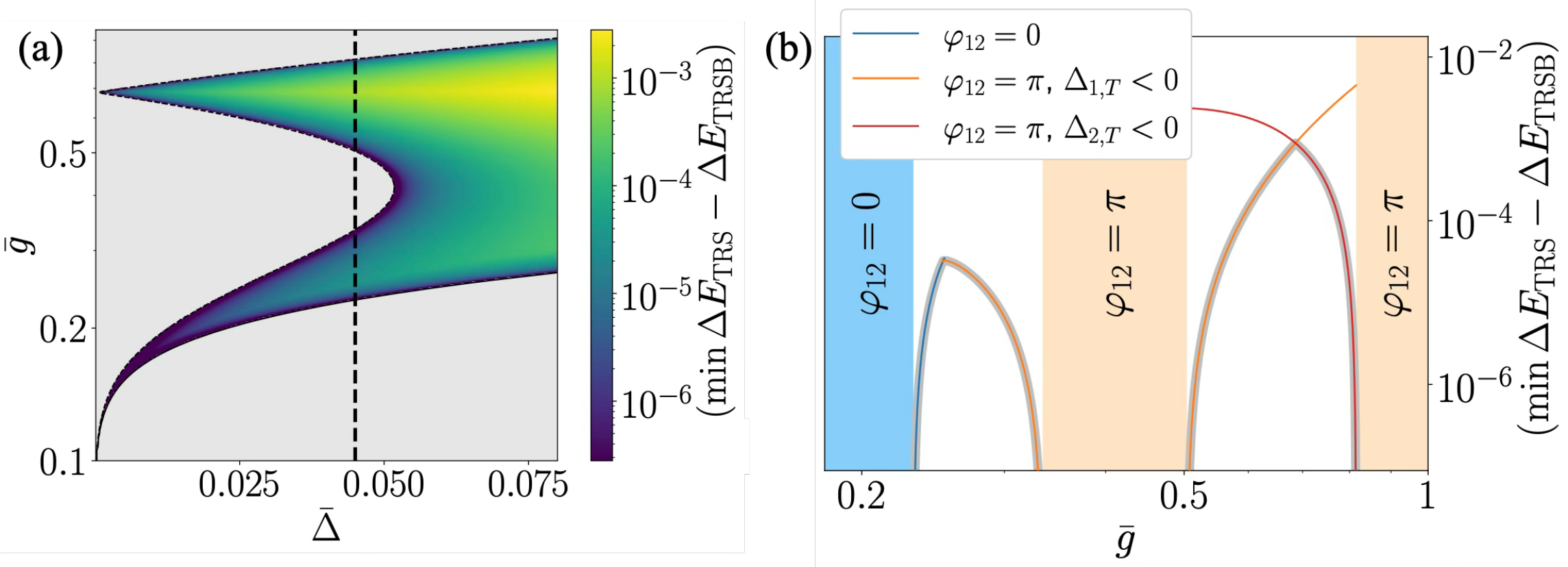}
    \caption{
    (a) Energy difference, with $U$ set equal to 1, between the TRS state with the lowest energy and the TRSB state
        as a function of $\bar\Delta$ and $\bar g$ when $r_{12}=0.6$ and $g_1/g_2=2.0$.
    (b) Vertical line cut of panel (a) for $\bar\Delta = 0.045$. 
        In the region where the TRSB state is favored, the different colors denote
        which TRS state has the lowest energy.
    }
  \label{fig.deltaE}
\end{figure}

Figure~\ref{fig.deltaE}~(a) shows the difference between the value of $\Delta E$ for 
the TRS state for which $\Delta E$ is the smallest, ${\rm min}\Delta E_{\rm TRS}$,
and $\Delta E$ for the TRSB state for the same set of parameter values used 
for Fig.~\ref{fig.results_asym}. The results confirm that whenever Eqs.~\ceq{eq.gap}
admit a TRSB solution, that solution is also the one for which the energy is lowest.
Figure~\ref{fig.deltaE}~(b) shows the evolution of $[{\rm min}\Delta E_{\rm TRS}-\Delta E_{\rm TRSB}]$
along the vertical cut of Fig.~\ref{fig.deltaE}~(a) denoted by the dashed line.
We see that the energy difference grows rapidly close to the boundary between
a TRS state and a TRSB one, but does not show discontinuities, indicating that the
transition between TRS and TRSB phases is continuous.
In panel (b) the color
of the line denotes which TRS solution has the lowest energy.
In the lower TRSB interval this is the $\varphi_{12}=0$ solution for
$0.23\lesssim\bar g\lesssim 0.25$ (blue line), and the $\varphi_{12}=\pi$ solution with
$\Delta_{1,T}<0$, i.e., $(\varphi_1,\varphi_2)=(\pi,0)$, for $0.25\lesssim\bar g\lesssim 0.33$
(orange line). In the upper interval the competing TRS solutions are the two
$\varphi_{12}=\pi$ solutions, with $\Delta_{1,T}<0$ (orange line) and
$\Delta_{2,T}<0$ (red line).
In each interval the energy gain is largest at the point of maximal frustration,
where the two lowest TRS solutions cross and are degenerate, and where
$\min\Delta E_{\rm TRS}$, and therefore the energy gain, has a cusp.
The cusps do not signal a transition: $\Delta E_{\rm TRSB}$ is smooth across them,
and neither of the crossing TRS solutions is metastable.
At a TRS stationary point the mixed derivatives of $\Delta E$ with respect to the moduli
and the phases vanish, and the Hessian of $U\Delta E$ with respect to $(\varphi_1,\varphi_2)$
has positive diagonal elements, $2g_j\hat\chi_j\Delta_{j,T}^2$, and determinant
$4g_1g_2\hat\chi_1\hat\chi_2\Delta_{1,T}^2\Delta_{2,T}^2(1-g_1\hat\chi_1g_2\hat\chi_2)$.
A TRS solution is therefore stable against a rotation of the phases if
$g_1\hat\chi_1g_2\hat\chi_2<1$, and it is a saddle point of $\Delta E$ if
$g_1\hat\chi_1g_2\hat\chi_2>1$. The marginal case $g_1\hat\chi_1g_2\hat\chi_2=1$
corresponds to the softening of the Leggett mode.
In the TRSB regions of Fig.~\ref{fig.deltaE}~(a) all the TRS solutions have
$g_1\hat\chi_1g_2\hat\chi_2>1$, whereas outside these regions the lowest TRS solution has
$g_1\hat\chi_1g_2\hat\chi_2<1$.
The cusps therefore mark the points where, if the gaps were constrained to be real,
the system would undergo a first-order transition between two TRS states.
The TRSB state preempts this transition: the phases $\varphi_1$, $\varphi_2$, measured from
the induced gaps, rotate continuously from the TRS configuration on one side of the TRSB
interval to the one on the other side, which are the two TRS solutions that cross at
the cusp, with continuous transitions at the boundaries of the interval.

The analysis above shows that to
obtain a TRSB superconducting state in a two-band metal proximitized by an $s$-wave superconductor,
the bands in the metal must satisfy the following two general conditions:
(i) the dominant components of the zero-momentum pairing susceptibility must be intraband; 
(ii) the interband pair scattering must exceed the intraband one.
These conditions allow us to identify which superconducting heterostructures
can support a TRSB superconducting state even when no magnetic field or ferromagnetism is present.

Let us consider Al/InAs heterostructures~\cite{shabani2016,wickramasinghe2018a,mayer2019,pan2025}.
These systems are of great interest because of the large SOC and $g$-factor
of InAs, which make them ideal for the realization of topological superconducting 
states~\cite{lutchyn2018,flensberg2021,schiela2024,pita-vidal2026}.
The 2DEG in the InAs layer is described by the Hamiltonian $H(\kk)=\xi(\kk)\sigma_0+\bb(\kk)\cdot\ssigma$, 
where $\{\sigma_i\}$ are the $2\times2$ Pauli matrices in spin space and $\bb(\kk)$ is the spin-orbit field.
The helical bands satisfy condition (i) when 
the spin-orbit splitting is much larger than the gaps.
On the other hand, these bands have the same orbital wavefunction and differ only in their spin
texture. As a consequence, the even-parity part of any spin-independent interaction gives
equal intraband and interband pair scattering, and condition (ii) can only be satisfied by
its odd-parity part. The configuration in which the gaps of the two helical bands have
opposite sign is a helical $p$-wave state, and condition (ii) requires the $p$-wave
component of the interaction to be attractive and, by Eq.~\ceq{eq.det}, strong enough for the
2DEG to be close to an intrinsic helical $p$-wave instability. 
This is an unlikely scenario, and so we conclude that a TRSB superconducting state is
improbable in heterostructures formed by an $s$-wave superconductor placed in proximity
to a 2DEG, even when the bands are spin-split by a strong SOC.
This is consistent with the experimental observation that in Josephson junctions
based on Al/InAs heterostructures no JDE is observed unless a magnetic field is
present~\cite{baumgartner2022,costa2023,lotfizadeh2024,yu2026}.
\begin{figure}[htb]
    \includegraphics[width=0.8\linewidth]{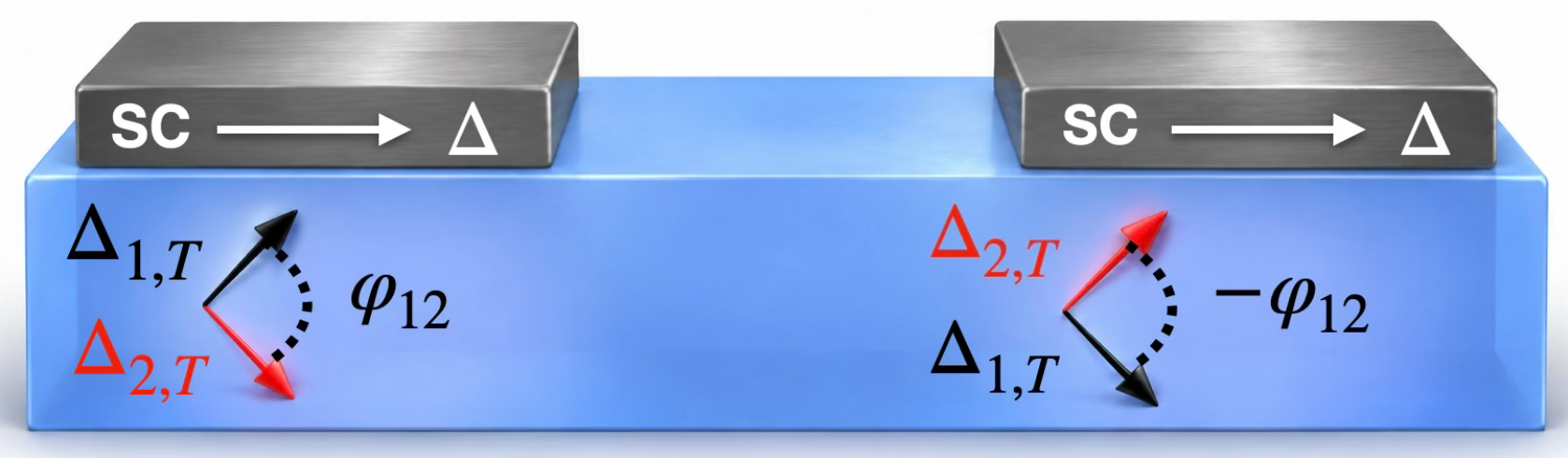}
    \caption{Sketch of Josephson junctions formed by superconducting leads 
             in opposite TRSB states.}
    \label{fig.sketch_JJ}
\end{figure}

The situation is different for heterostructures formed by an $s$-wave superconductor
placed on a 3D Dirac semimetal such as \cdas.
For this case, we can identify bands 1 and 2 as the surface and bulk bands of the semimetal;
the two bands are distinguished by their spatial structure: surface and bulk
states have different profiles, so that already the even-parity, short-range part of the
interaction gives an interband pair scattering different from the intraband ones.
Condition (ii) can then be satisfied without any odd-parity attraction.
As a consequence, heterostructures formed by an $s$-wave superconductor and a 3D Dirac semimetal
are more likely to realize a spontaneously TRSB superconducting state. 
This is supported by several experimental
results for Josephson junctions and SQUIDs based on heterostructures formed
by Al and the semimetal \cdas:
(i)   the observation of soft Leggett modes~\cite{cuozzo2024c};
(ii)  the observation of half-integer Shapiro steps in {\em single} Josephson junctions~\cite{yu2018e};
(iii) a zero-magnetic-field JDE in asymmetric SQUIDs~\cite{yu2024,travaglini2025}.
Each of these results is explained by assuming the establishment of a TRSB canted superconducting
state in the Al/\cdas heterostructure with a relative angle $\varphi_{12}$ between the pairing gaps
for the surface and bulk bands approximately equal to $\pm\pi/2$,
with the leads of the Josephson junction picking opposite states
of the time-reversed pair as schematically shown in Fig.~\ref{fig.sketch_JJ}.
When the leads of the Josephson junction have opposite values of $\varphi_{12}\neq 0,\pi$,
in addition to TRS, also the left-right mirror symmetry is broken. 
The combined breaking of TRS and mirror symmetry is what can explain in a natural way the 
observation of a field-free JDE in Ref.~\cite{yu2024,travaglini2025}, and is the most compelling explanation
for the presence of half-integer Shapiro steps in {\em single} Josephson junctions
measured in Ref.~\cite{yu2018e} and for the structure of the Shapiro steps observed in SQUIDs
arising from a soft Leggett mode~\cite{cuozzo2024c}.

The experimental observation of a zero-magnetic-field Josephson diode effect
in junctions based on heterostructures formed by an $s$-wave superconductor and a 3D Dirac semimetal~\cite{yu2024}
has been puzzling given the absence of any magnetic ordering in these systems.
We have presented a microscopic model that gives a compelling mechanism
for the establishment in such heterostructures of a multiband superconducting
ground state that spontaneously breaks time-reversal symmetry,
a necessary condition for the presence of a Josephson diode effect. 
The mechanism relies on the frustration resulting from the tendency 
of the pairing fields of the two bands in the substrate to align
with the phase of the $s$-wave superconductor and the tendency to anti-align due
to a repulsive interband pair interaction.
We have identified the general conditions for the realization of the 
TRSB superconducting state, and determined in which regions of the parameter
space such a state is favored. We have found that the resulting phase diagram can exhibit interesting
and surprising reentrant behavior whenever the two bands are dissimilar,
for instance by having different magnitudes for the superconducting gap induced
via the proximity effect.
The results clarify when a TRSB state should be present in superconducting 
heterostructures and are therefore valuable for the realization
of field-free, and magnetization-free, nonreciprocal superconducting transport effects,
such as the superconducting diode effect and the Josephson diode effect,
that can be used to obtain compact superconducting diodes, devices that
can perfectly isolate qubits and quantum sensors from readout noise.


\begin{acknowledgments}
This work has been funded by the U.S. Department of Energy, Office of Basic Energy Sciences, via Award No.
DE-SC0022245. 
\end{acknowledgments}


\bibliography{canted_sc}



\onecolumngrid
\newpage
\makeatletter

\begin{center}
\textbf{\large Supplementary Materials for: ``\@title ''} \\[10pt]
Jonathan Schirmer$^{1}$, Wei Pan$^{2}$, Javad Shabani$^{3}$, and Enrico Rossi$^{1}$ \\[4pt]
\textit{$^1$Department of Physics, William \& Mary, Williamsburg, Virginia 23187, USA}\\
\textit{$^2$Sandia National Laboratories, Livermore, California 94551, USA}\\
\textit{$^3$Center for Quantum Information Physics, New York University, New York, New York 10003, USA}\\
\end{center}
\vspace{10pt}

\setcounter{page}{1}  
\setcounter{figure}{0}
\setcounter{section}{0}
\setcounter{equation}{0}

\renewcommand{\thefigure}{S\@arabic\c@figure}
\makeatother

\appendix
\counterwithout{equation}{section}               
\renewcommand{\theequation}{S\arabic{equation}}  

\setlength{\parindent}{0pt}
\setlength{\parskip}{0.5\baselineskip}

Below we provide a few additional details on the conditions required to realize
a TRSB, canted, superconducting state, and additional results for the phase diagram
for asymmetric cases.


\section{The ``loop condition''}

The canted, TRSB, superconducting state described in the main text requires
at least three separate superconducting pairings. This allows the identification
of a physical, gauge invariant, phase difference between two of the gaps, $\varphi_{12}$.
In our case, one pairing is the one of the parent superconductor, and the other
two are the pairings of the two bands in the metal, which are independent
when condition (i) of the main text is satisfied.
In addition, the couplings among the three pairings must be frustrated.
This leads to the ``loop condition'' discussed in this section, whereas
condition (ii) of the main text is discussed in the next section.

\begin{figure}[h!!!!!!!b]
 \includegraphics[width=0.6\linewidth]{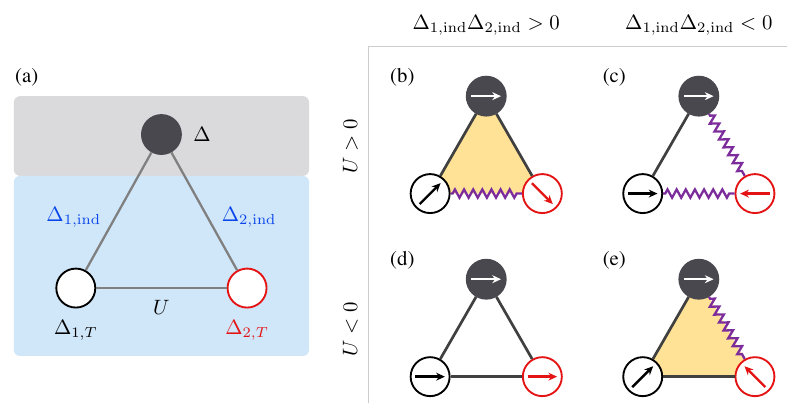}
    \caption{The loop condition, Eq.~\ceq{eq.loop}.
    (a)~The pairing field $\Delta$ of the parent superconductor (gray) and the total
    pairing gaps $\Delta_{1,T}$, $\Delta_{2,T}$ of the two bands of the metal (blue) form a
    closed loop with three links: the induced gaps $\Delta_{1,\rm ind}$, $\Delta_{2,\rm ind}$,
    and the interband interaction $U$.
    \mbox{(b)--(e)}~The four sign combinations with $\Delta_{1,\rm ind}>0$: same-sign induced gaps
    (left column, e.g., $s$-wave parent) and opposite-sign induced gaps (right column,
    e.g., $s_\pm$ or $d$-wave parent), for repulsive (top row) and attractive (bottom row) $U$.
    Straight (zigzag) lines are sign-preserving (sign-reversing) links, which favor a
    relative phase $0$ ($\pi$) between the pairings they connect; a link is sign-reversing
    if $\Delta_{j,\rm ind}<0$ or $U>0$.
    Arrows show the phases of $\Delta$ (white), $\Delta_{1,T}$ (black), and $\Delta_{2,T}$ (red).
    In (c) and (d) the number of sign-reversing links is even, and the TRS configurations
    $(\varphi_1,\varphi_2)=(0,\pi)$ and $(0,0)$, respectively, satisfy all links.
    In (b) and (e) the number of sign-reversing links is odd, $U\Delta_{1,\rm ind}\Delta_{2,\rm ind}>0$: the loop is
    frustrated (shaded yellow) and the phases can cant into one of two time-reversed TRSB
    configurations, one of which is shown.}
  \label{fig.loop}
\end{figure}

Using the notation of the main text, since $\hat\chi_1>0$ and $g_1=UN_1(0)$,
the condition~\ceq{eq.comp} requires
$\sgn(U)=\sgn(\Delta_{2,\rm ind}/\Delta_{1,\rm ind})$, and this can be cast as
the following inequality
\beq
 U\,\Delta_{1,\rm ind}\,\Delta_{2,\rm ind} > 0 .
 \label{eq.loop}
\enq
The requirement $U>0$ in Eq.~\ceq{eq.Hint} is the special case of equal-sign
induced gaps.  Frustration is a property of the closed loop
parent~$\to$~band~1~$\to$~band~2~$\to$~parent, present when the number of
sign-reversing links is odd, counting a repulsive $U$ and each negative induced
gap as one such link (Fig.~\ref{fig.loop}).  An intrinsic two-band superconductor has a single link
and no loop, which is why intrinsic frustration needs at least three
components; in our setting the parent supplies the two remaining legs.
Equation~\ceq{eq.loop} shows that a parent seeding \emph{opposite-sign} gaps on
the two bands inverts the required sign of $U$.  This happens for a
sign-changing parent ($s_\pm$ or $d$-wave) when the two bands sample
opposite-sign regions of the parent gap.


\section{Effect of intraband interactions}
\label{sm.intra}
Restoring intraband couplings $V_{jj}=U_j$, with $g^{(d)}_j\equiv U_jN_j(0)$,
the degeneracy condition becomes
\beq
 \big(1+g^{(d)}_1\hat\chi_1\big)\big(1+g^{(d)}_2\hat\chi_2\big)
   = g_1\hat\chi_1 g_2\hat\chi_2 ,
 \label{eq.intra}
\enq
which replaces Eq.~\ceq{eq.det}.
For repulsive intraband couplings, $g^{(d)}_j\ge 0$, the left-hand side of Eq.~\ceq{eq.intra}
is larger than $g^{(d)}_1\hat\chi_1g^{(d)}_2\hat\chi_2$, so the degeneracy condition can be
satisfied only if $g_1g_2>g^{(d)}_1g^{(d)}_2$, i.e., $U^2>U_1U_2$: the interband pair
scattering must exceed the geometric mean of the intraband ones.
This is the precise form of condition (ii) of the main text.
In the symmetric case Eq.~\ceq{eq.intra} becomes
$\hat\chi(g-g^{(d)})=1$ in place of $\hat\chi g=1$, i.e., the threshold is
shifted by exactly the intraband coupling,
\beq
 g_c \longrightarrow g_c + g^{(d)} .
 \label{eq.shift}
\enq

\section{Additional phase diagrams for asymmetric cases}
\label{sm.asym}
\begin{figure}[h]
 \includegraphics[width=0.75\linewidth]{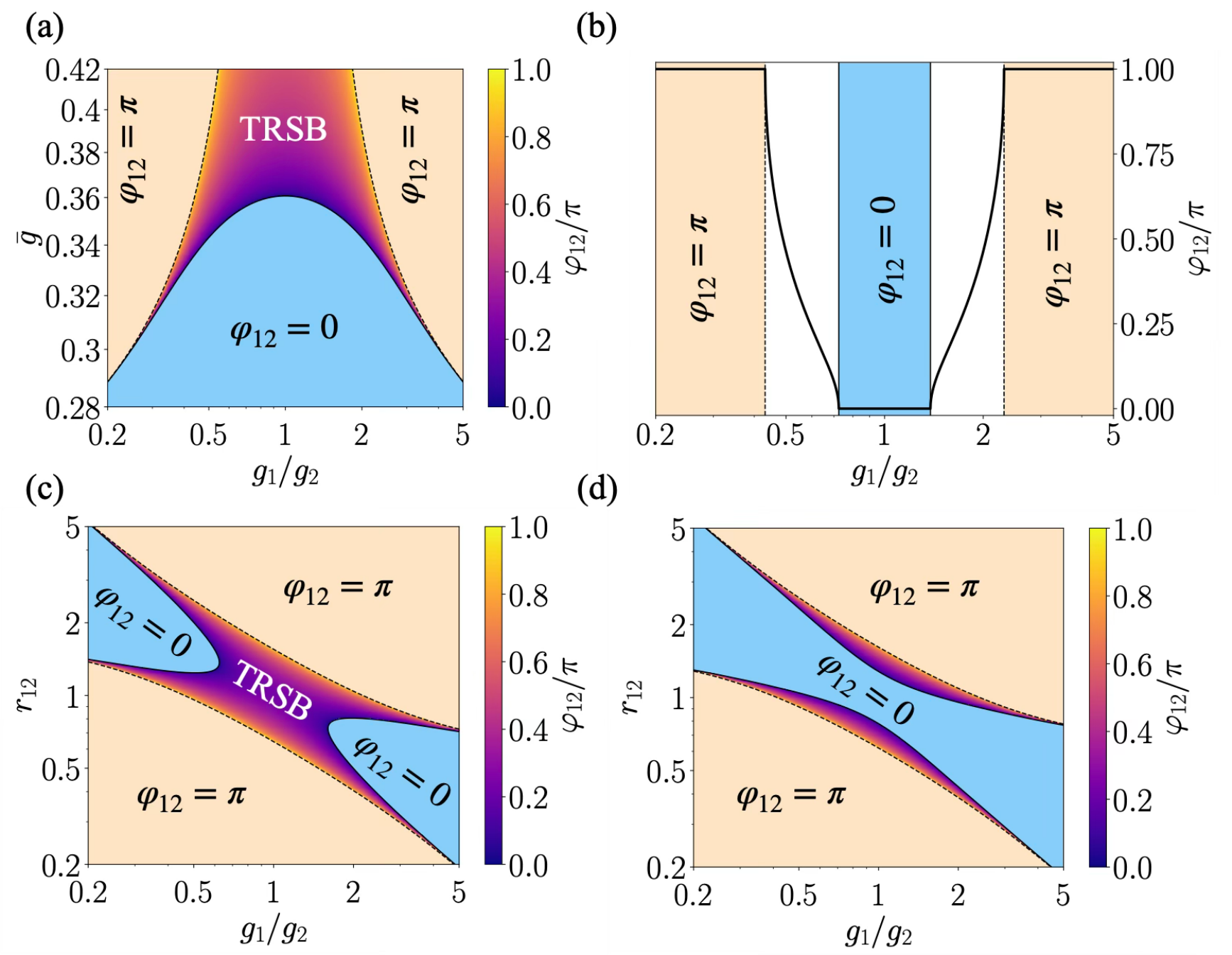}
    \caption{
    Phase diagrams and line cut for asymmetric cases: $r_{12}\neq 1$ and/or $g_1/g_2\neq 1$.
    (a) Phase diagram in ($g_1/g_2$,$\bar g$) space when $r_{12}=1$ and $\bar\Delta=0.25$.
    (b) Horizontal line cut of panel (a) for $\bar g = 0.357$.
    (c) Phase diagram in ($g_1/g_2$,$r_{12}$) space when $\bar\Delta=0.25$ and $\bar g = 0.37$.
    (d) Phase diagram in ($g_1/g_2$,$r_{12}$) space when $\bar\Delta=0.25$ and $\bar g = 0.35$.
    }
  \label{fig.sm_asym}
\end{figure}

Figure~\ref{fig.sm_asym}~(a) shows the phase diagram, as a function
of the ratio $g_1/g_2$ and $\bar g$, when $r_{12}=1$ and $\bar\Delta=0.25$.
Also in this case we can observe reentrant behavior.
The log-scale for the $x$-axis
makes evident that the phase diagram is symmetric under $g_1/g_2\to g_2/g_1$.
Figure~\ref{fig.sm_asym}~(b) is a horizontal line cut of Fig.~\ref{fig.sm_asym}~(a)
for $\bar g = 0.357$.
Figures~\ref{fig.sm_asym}~(c),~(d) show the phase diagram
in the $(g_1/g_2, r_{12})$ parameter space for $\bar\Delta=0.25$ and
two different values of $\bar g$, $\bar g = 0.37$ and $\bar g = 0.35$, respectively.
The difference between the two
panels illustrates the fact that as $\bar g$ decreases the phase space of
the TRSB state is reduced.

\end{document}